# Magnonic Magnetoelectric Coupling in Ferroelectric/Ferromagnetic Composites

*Levan Chotorlishvili, Chenglong Jia, Diana A. Rata, Liane Brandt, Georg Woltersdorf, and Jamal Berakdar\**

Composite materials consisting of coupled magnetic and ferroelectric layers hold the promise for new emergent properties such as controlling magnetism with electric fields. Obviously, the interfacial coupling mechanism plays a crucial role and its understanding is the key for exploiting this material class for technological applications. This short review is focused on the magnonic-based magneto-electric coupling that forms at the interface of a metallic ferromagnet with a ferroelectric insulator. After analyzing the physics behind this coupling, the implication for the magnetic, transport, and optical properties of these composite materials is discussed. Furthermore, examples for the functionality of such interfaces are illustrated by the electric field controlled transport through ferroelectric/ferromagnetic tunnel junctions, the electrically and magnetically controlled optical properties, and the generation of electromagnon solitons for the use as reliable information carriers.

## 1. Introduction

Multiferroics (MFs) are materials that exhibit more than one order parameter that may be coupled to each other.[1,2] In addition to the preexisting properties, the coupling between the order parameters result in emergent features that can be exploited for applications such as four-point memory devices. Fueled by these prospects and armed with advances in material synthesis and device fabrication and characterization, intense research in recent years has established solid fundamental understanding

Dr. L. Chotorlishvili, Dr. D. A. Rata, L. Brandt, Prof. G. Woltersdorf, Prof. J. Berakdar
Institut für Physik
Martin-Luther Universität Halle-Wittenberg
06099 Halle/Saale, Germany
E-mail: jamal.berakdar@physik.uni-halle.de

Prof. C. Jia
Key Laboratory for Magnetism and Magnetic Materials
Ministry of Education
Institute of Theoretical Physics
Lanzhou University
73000 Lanzhou, China

The ORCID identification number(s) for the author(s) of this article can be found under https://doi.org/10.1002/pssb.201900750.

of the electronic, magnetic, and optical properties of these compounds. Exhaustive overviews on different aspects of the field were given in a number of articles such as refs. [1,2]. Our goal here is more limited with the specific goal to understand the magnetic dynamic and static properties at the interface of two-phase composite MFs. Specifically, of interest are oxide-based magnetoelectrics (MEs) consisting of two parts: one part is ferromagnetic (FM) interfaced with another ferroelectric (FE) part, for instance, Co/BaTiO$_3$. This class of MF holds the promise for new applications in magnetism, spintronics, and magnetooptics. As discussed here, for instance, the local noncollinear magnetic order formed at the coupled interfaces can be steered by electric fields and results in an emergent local spin–orbital coupling (SOC) of traversing charge carriers. This effect can be useful for electrically controlled magnetic tunnel junctions. In addition, the magnetic permeability turns electrically sensitive which is relevance for applications in magnetophotonics. Compounds that exhibit within the same material coupled ferroelectric and magnetic (possibly spiral) ordering are well documented. Prototypical examples are BiFeO$_3$ and YMnO$_3$. A substantial body of literature on MF already exists (for instance refs. [1–24] and references therein). Also, various aspects of the ME coupling mechanism are well understood,[25–33] down to the microscopic level. How ME operates at FE/FM interfaces and which emergent material properties are brought about by this coupling were the subject of numerous investigations.[34–43] Generally, for FM/FE bilayer local and nonlocal electronic correlations and kinetic exchange effects at the interface may change strongly the electronic and magnetic properties at the atomic constituents at the interface (see, e.g., ref. [19]) and even quench the interfacial local magnetic moments. In addition, these interactions lead to a change in the elastic properties on the FE side due to interfacial strain caused by interfacial intercalation.[13–16] While these electronic and elastic processes result in total in a lowering of the total free energy of the system, and therefore in an interfacial ME coupling, the local interfacial magnetism is not in equilibrium with the bulk magnetization in the FM. As discussed later, on top of the previous electronic and elastic mechanism the energy can therefore be lowered further, albeit at a smaller energy scale, by forming in general a spiral ordering extending from the interfacial (possibly quenched) magnetic order out to the bulk one.[44] This type of transversal magnetic dynamics can be viewed as a superposition of frozen magnons. While energetically the contribution of this mechanism relative to the electronic part might be small, its influence on the interfacial spintronic properties can be very important. For example, as well established a noncollinearity in the magnetic order leads to the emergence of a gauge field that couples the carrier spin to its orbital dynamics. The strength of this coupling is related to the period of the spiral. As the spiral in FM/FE junction is confined to the interface in region limited by the spin-diffusion length, the spiral period can be relatively small indicating a relatively strong SOC. In addition, the direction of the spiral is linked to the FE order that causes the spiral in the first place, allowing thus a control of the spiral by electric means. The rest of this article will be indeed focused on this type of coupling and the consequences thereof for the physical properties of FE/FM layers.

The article is organized as follows: in Section 2, we discuss the fundamentals behind the formation of the noncollinear spin order at FE/FM interfaces. In Section 3, we analyze the ME coupling in epitaxial lanthanum strontium manganite (LSMO)/lead zirconate titanate (PZT) samples, and in Section 4 we study the possibility of FE control of anisotropic damping in MF tunnel junctions. Section 5 is devoted to the effect of spiral magnetic order on propagating electromagnetic (EM) waves. Conclusions are made in Section 6.

## 2. Noncollinear Spin Order at FE/FM Interfaces and Tunnel Junctions

### 2.1. Theory of the Magnonic-Based Interfacial ME Coupling

Experimental and theoretical studies on FE layers in contact with ferromagnets evidence that electronic reconstruction[12] may occur as a result from a hybridization of the orbitals at the interfacial layers.[12–15] The electronic rearrangement leads to atomic reconstruction accompanied by surface stress and also a change, or even a quenching of the magnetic and FE properties at the interfacial layers.[12–19] These processes are local; away from the interface the bulk properties of FM and FE materials can be markedly different from those in the transitional region in the vicinity of the interface between both materials. Topological interfacial distortions, such as dislocation lines or a local noncollinear topological order, may, however, survive even in the bulk region. Other excitations due to the FE/FM coupling can be atomistic localized to the interface. For example, normal (sp-bonded) metal reacts to a local dielectric polarization (for instance, at the interface) by an electrostatic screening. The screening charge is tight to the electrostatic distortion on a scale set by the charge screening length in the metal, which is typically on the order of a few angstroms.[12] This type of atomically sharp interfacial screening can be accounted by self-consistent field calculations[12] performed in a relaxed geometry for sufficiently large supercell. If the metal is FM, the local screening charge is also polarized resulting in a changed local magnetic moments, meaning longitudinal variation of spin order. The transversal magnetic order at the interface needs then to be relayed to the bulk magnetization in the FM, a process that occurs on markedly different length scale than the interfacial charge reconstruction. This is indeed consistent with the respectively different energy

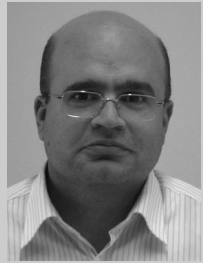

**Levan Chotorlishvili** studied physics at Tbilisi State University (Georgia) and received a Ph.D. in theoretical and mathematical physics in 1998 and habilitation in 2005. He worked as a theoretical physicist at Tbilisi State University from 2000 to 2008, the University of Augsburg from 2008 to 2010, and the University of Heidelberg in 2011. From the 2012 he is working as scientist at the Martin Luther University, Halle-Wittenberg, Germany. His research interests include spintronics and spin-caloritronics, KAM theory and quantum chaos, quantum information theory, ultrahigh energy neutrino physics, open quantum systems, and quantum measurements.

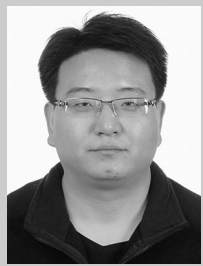

**Chenglong Jia** is a Professor at the School of Physical Science and Technology at Lanzhou University, China. Before joining the faculty of Lanzhou University, he was working at Martin Luther University, Halle-Wittenberg, Halle, Germany and joined the SFB762 as a DFG-Mercator fellow after accepting the appointment in Lanzhou. His main research interests include correlated electron systems and quantum magnetism.

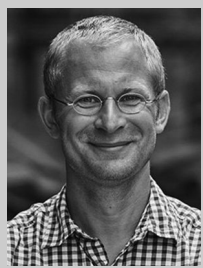

**Georg Woltersdorf** graduated from Simon Faser University (Canada) with a Ph.D. in physics in 2004. Following a postdoc appointment at the University of Regensburg (Germany), he worked as a research group leader and staff scientist at University of Regensburg. In 2013, he was appointed as a full professor at the Martin Luther University, Halle, Germany. His main research interests include spin dynamics, spin currents, and ultrafast phenomena in magnetic heterostructures and devices.

scales implying correspondingly different wave lengths for the involved excitations. Furthermore, being a vector field, an interfacial discontinuity of the magnetization can extend deep into the bulk. Evidence of these statements was provided, for example, by ferromagnetic resonance (FMR) experiments of polycrystalline, 5–40 nm-thick hcp Co-layers deposited on 15 nm-thick tetragonal $BaTiO_3$ substrate (BTO). A pronounced influence of the FE polarization of BTO[45–47] on the spin-wave excitations in the Co layer was observed. From the induced anisotropy in the FMR, a related ME coupling strength of $0.27 \, s \, F^{-1}$ was deduced. As concluded also theoretically, only a linear ME coupling can cause the observed asymmetry in the angular-dependent FMR spectra.[45–47] These experiments demonstrated also the possibility of electric field tuning of FMR in a coupled FM/FE layered system. How is it possible to describe theoretically the interfacial FE/FM coupling and quantify its length and energy scale? For an answer, let us focus on the FM part and perform $s$–$d$-type analysis to find the equilibrium state of the delocalized spin-dependent carriers ($s$-type) to the localized ($d$) moments. The induced spin imbalance is described by the spin-density operator $\hat{s} = \sum_{\sigma\sigma'} \psi_\sigma^+(r)\hat{\sigma}_{\sigma\sigma'}\psi_{\sigma'}(r)$. Here, the field operators $\psi_\sigma^+(r)$ and $\psi_\sigma(r)$ create (annihilate) charge carriers with spin $\frac{\hbar}{2}\sigma$. We are interested in the dynamics and stationary states of the induced spin imbalance. The relevant Hamiltonian within the $s$–$d$ coupling reads

$$H = \frac{\hbar^2}{2m} \sum_\sigma \int d\vec{r}\, \vec{\nabla}\, \psi_\sigma^+(\vec{r}) \cdot \vec{\nabla}\psi_\sigma(\vec{r}) + \int d\vec{r}[V(\vec{r})\hat{n}(\vec{r}) + H_{sd}] \tag{1}$$

In addition to the kinetic part, $H$ accounts for charge scattering ($\hat{n}(\vec{r}) = -e\sum_\sigma \psi_\sigma^+(\vec{r})\psi_\sigma(\vec{r})$) from the interface potential $V$ and from the magnetic background, namely, $H_{sd} = J_{ex}\hat{s}\cdot\vec{e}_M$.

Quantum fluctuations of the magnetization $\vec{M}$ are assumed to be suppressed (e.g., due to magnetic anisotropy and finite temperatures) such that capturing the dynamics of $\vec{M}$ amounts to describing the classical unit vector field $\vec{e}_M = \vec{M}/M_s$ (where $M_s$ is the saturation magnetization). The strength of the coupling of $\hat{s}$ to the classical moments $\vec{e}_M$ is set by the exchange coupling $J_{ex}$. The dynamics of the itinerant spin density $\vec{s} = \langle\hat{s}\rangle$ is governed by Bloch equation

$$\frac{d\vec{s}}{dt} + \vec{\nabla}\Theta = -\frac{1}{\tau_{ex}}\vec{s} \times \vec{e}_M - \frac{\vec{s}}{\tau_{sf}} \tag{2}$$

Here, $\Theta = \frac{\hbar}{2m}\langle \mathrm{Im}[\psi^+\hat{\sigma}\otimes\vec{\nabla}\psi]\rangle$ is the spin current density, $\tau_{ex} = \hbar/(2J_{ex})$, and $\tau_{sf}$ is the spin-flip relaxation time due to scattering from impurities. For a full analysis of Equation (2), we refer to ref. [44]. The solution of Equation (2) yields (with respect to $\vec{e}_M$) the transversal and longitudinal components

$$s_\parallel = C_\parallel e^{-z/\lambda_m}$$
$$s_\perp^x + is_\perp^y \approx C_\perp e^{-(1-i)\hat{Q}_m\vec{r} + ia_0} \tag{3}$$

with $\vec{Q}_m = \frac{1}{\sqrt{2}D_0}[0,0,1]$. The effective spin-diffusion length is $\lambda_m = \sqrt{D_0\tau_{sf}}$ with the diffusion coefficient $D_0$, which is material dependent and usually on the order of a few nanometers. The coefficients $C_\parallel$ and $C_\perp$ follow from the overall electric neutrality condition $C_\parallel = \eta P_s/\lambda_m$, $C_\perp = (1-\lambda_m)Q_m P_s$, where $\eta$ is the spin polarization. The spin-dependent carrier-density accumulation $P_s$ is not in equilibrium with $\vec{M}$ and therefore acts with a torque on it leading to a twist in the interfacial magnetization. The resulting modulations are

$$\Delta\vec{M}_\parallel = \eta\frac{\vec{P}_s}{d}\mu_B, \quad \Delta\vec{M}_\perp = (1-\eta)\frac{\vec{P}_s}{d}\mu_B \tag{4}$$

where $d$ is the thickness of the FM film. Equation (3) and (4) indicate the formation of an interfacial spiral that relaxes to the uniform bulk magnetization. Viewed from the perspective of the bulk FM, the spiral constitutes a superposition of standing spin-wave excitations of the FM (frozen magnons) localized near the FM/FE interface. The modulations disappear for vanishing FE polarization. Hence, the spiral is indeed a manifestation of

the ME coupling which can be identified to be of a linear type having the form[44]

$$E_{ME} = \alpha_{ME}P_0^z M_0, \quad \alpha_{ME} = \frac{\eta J_{ex}}{ed M_s} \tag{5}$$

We stress that we started from the electronically and structurally relaxed state (meaning the state with the lowest energy for the FM/FE system) and relaxed on top of that spin ordering. This hierarchal doing is justified because the electronic and the phononic excitations have a significantly higher energy than the magnonic (i.e., transversal spin) excitations. This means, in the system which we are dealing with many other ME coupling terms due to interfacial charge migration, intercalation, and magnetoelastic coupling maybe operational. We assume them here to be residual interaction and are not abolished by the magnonic coupling (due to energy consideration). Equation (5) is an expression of the magnonic contribution to the ME coupling energy. Depending on material compositions and intrinsic material properties, the various contributions to the ME coupling might be different. For instance, in ref. [44] using the appropriate choice of the layered structure and buffer layers, the elastic contribution was varied. The magnonic contribution can be assessed by comparing materials with different spin diffusion lengths, as one can see from Equation (3). For instance, the spin diffusion length is $8.5 \pm 1.5$ nm in Fe and $38 \pm 12$ nm in Co.[48,49] For LSMO, depending on doping the spin diffusion length can be as short as few Angstroms and therefore the magnonic contribution (e.g., for PZT/LSMO interfaces[50]) is expected to be less prominent. As clear from the aforementioned derivation, the electronic interaction is the principal source of the various ME couplings. The magnonic contribution may be operational without affecting the electrostatic charge distribution (on the contrary, ultimately the magnonic part is caused by local electron correlations and their interplay by exchange interactions, as clear from the aforementioned derivation). Due to the small energy scale for the magnonic modes, a weak interaction may be sufficient to trigger magnonic excitations. For example, if the FM and the FE part are separated by a semiconducting or insulating layer, the dielectric screening of the FE polarization on the metallic FM side may trigger the interfacial magnons that are coupled to the FE polarization.

## 2.2. Electric Tuning of FM Resonance by ME Coupling

How can the magnonic ME coupling contribution be detected experimentally?

To answer this question, two observations are relevant: 1) The interfacial noncollinearity in the localized moments implies a SOC for carriers in the FM traversing the noncollinear region.[51] This type of SOC might be useful for interfacial, electric field-controlled spintronic applications. 2) The magnonic modes of the FM part (decoupled from the FE part) are altered when interfaced with the FE part. This change can be traced via FM resonance[52] which probes the magnetic response to a perpendicular strong static and a weak radio frequency (RF) magnetic field. Varying the amplitude of the static field for a fixed amplitude and the frequency $\omega$ of the RF field yields the spectrum of absorbed power which peaks at the FM resonance. FMR position is related to the effective magnetic field (stiffness). The FMR peak is usually broadened due to a number of internal magnonic relaxation (damping) mechanisms;[53] the above interfacial SOC will add to them.[24,49] It can be distinguished from others by tuning the FE polarization via an external electric field, as evidenced by recent experiments.[53,54] We note that FMR can also be performed locally and in a time-resolved manner which can be exploited to distinguish the internal characteristics of magnetic excitations.

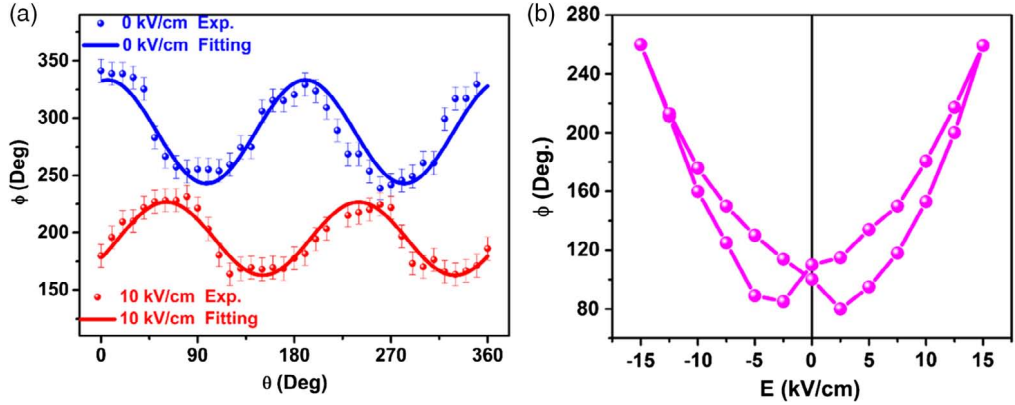

**Figure 1.** Measurements of FMR spectroscopy. a) Change in mixture phase $\phi$ with angle $\theta$ between the static magnetic field $H$ and magnetization easy axis under unpoled state and FE-polarized state by a normal electric field $E = 10\,\text{kV}\,\text{cm}^{-1}$ in Co/Pb(Mg$_{1/3}$Nb$_{2/3}$)$_{0.7}$Ti$_{0.3}$O$_3$ (Co/PMN-PT) heterostructure. Solid lines show fitting curves with introduced $C_{2v}$ symmetry. b) Change in mixture phase $\phi$ with electric field $E$ applied normally to the Co/PMN-PT.

Suppose we have a linear ME at FM/FE interface, the induced magnetization $\delta m_i$ by applying microwave field is

$$\delta m_i = \chi_{ii}^m h_i + \frac{\alpha_{ME}^{ji}}{\mu_0} e_j \tag{6}$$

where $\alpha_{ME}$ is the ME tensor; $\chi^m = \mu - 1$ is the intrinsic magnetic susceptibility in the FM subsystem; $h$ and $e$ are the AC magnetic field and electric field, respectively, satisfying the Maxwell–Faraday equation $\nabla \times e = -\frac{1}{\mu}\frac{\partial h}{\partial t}$, which reconstitutes the magnetic/electric field codriven *effective* dynamic magnetic susceptibility $\tilde{\chi}^m$, with

$$\text{Im}[\tilde{\chi}^m] = \beta\text{Im}[\chi^m]\cos\phi + \beta\text{Re}[\chi^m]\sin\phi \tag{7}$$

$$\text{Re}[\tilde{\chi}^m] = \beta\text{Re}[\chi^m]\cos\phi - \beta\text{Im}[\chi^m]\sin\phi \tag{8}$$

where $\text{Re}[\chi^m]$ and $\text{Im}[\chi^m]$ are the real and imaginary parts of the *intrinsic* magnetic susceptibility $\chi^m$ of FM material without ME interactions, respectively. The prefactor $\beta e^{i\phi} = (1 + \frac{\alpha c}{n})$ contains the phase $\phi$ between the induced magnetization and the applied ac magnetic field and is mainly due to the complex refractive index $n$ of the FM subsystem, but the magnitude is determined by the ratio $\alpha c/n$. For the case when linear ME interaction $\alpha = 0$, one has $\phi = 0$ and $\beta = 1$, which returns to the magnetic susceptibility in natural isolated FM systems. However, the comparable real and imaginary part of complex refractive index ($|n| = 10^3$) of typical FM metals (Fe, Co, and their alloys at 140 GHz)[55] and the large ME coupling ($\alpha = 10^{-7}s/m$) at the FM/FE interface could result in nontrivial phase $\phi$ and considerable modification of the effective magnetic susceptibility $\tilde{\chi}^m$ in the GHz range. In particular, considering that the real ($\text{Re}[\chi^m]$) and imaginary ($\text{Im}[\chi^m]$) parts of natural FM under microwave irradiation normally demonstrate themselves as a dispersive and an absorptive Lorentzian line shape with the frequency of the applied microwave field, such nontrivial phase $\phi$ would result in a mixture of the dispersive and absorptive line shapes of FM resonance, resulting in a transformable magnetic permeability $\tilde{\chi}^m$ provided that $\phi$ is ferroelectrically tunable (cf. **Figure 1**). This transformable behavior of the dynamic permeability is not only the *general* but also the *unique* feature of linear ME systems.[55]

## 3. ME Coupling in Epitaxial LSMO/PZT/LSMO Samples

The electronic ME coupling mechanism discussed earlier will be examined statically for a model sample system using a configuration where a magnetoelastic contribution is not expected.[56] The charge-mediated magnetoelectric coupling effect is driven by a buildup of an interfacial spiral spin density in FM/FE composites. Depending on the FE polarization state in a FM/FE composite, the surface spiral spin density induced in the FM layer changes its direction. Here, two identical strain states are compared (the FE polarization is for our experiments always oriented along the film normal). For the experimental geometry shown in **Figure 2**, this is expected to lead to an out-of-plane component of the magnetization in the vicinity of the FM/FE interface over the length scale of the spin diffusion length in the magnetic material. We developed an experiment that is sensitive to minute changes ($<10^{-6}$ rad) of the polarization state of reflected light from an individual MF capacitor. However, magnetooptic effects (interaction with the FM layer) and electrooptic effects (such as the Pockels effects in the FE layer) can contribute to a measured change in the polarization state.[57] We prepared MF capacitors FM/FE/FM (FE = PZT, FM = LSMO) using pulsed laser deposition on strontium titanate (STO) (001) substrates. The experimental configuration and the sample structure are shown in Figure 2.

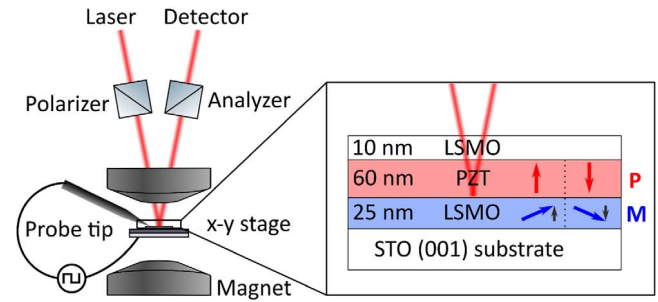

**Figure 2.** The left panel shows the experimental configuration of the experiments that study the static ME coupling in LSMT/PZT heterostructures. The polarization changes upon reflection are analyzed, whereas a voltage is applied to individual MF capacitors.

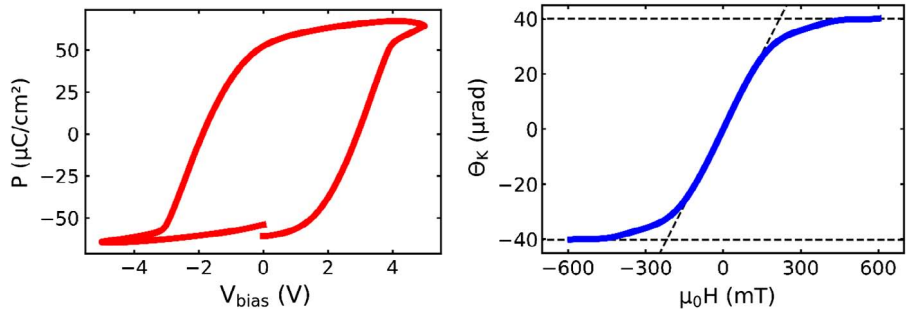

**Figure 3.** (Left) Electric polarization curve of the 60 nm-thick PZT layer measured by cycling the voltage on the FM/FE hybrid capacitor and integrating the current. (Right) Polar MOKE magnetization curve of the 25 nm-thick LSMO bottom electrode. The effective magnetization $\mu_0 M_{eff}$ is estimated to be 220 mT (dotted line).

First, the magnetic and static properties of the FE and FM layers of the heterostructure are characterized, as shown in **Figure 3**. In the FE PZT layer, a polarization of almost 60 µm cm$^{-2}$ is obtained. Even cycling the FE hysteresis more than $10^8$ times does not result in a sizable decay of the FE polarization. The magnitude of the saturation magnetization in the FM bottom layer (25 nm LSMO) at 300 K can be estimated from the polar magnetooptic Kerr effect (MOKE) loop. The magnetic saturation field indicates an effective magnetization $\mu_0 M_{eff} =$ 220 mT. SQUID measurements indicate that in our films $\mu_0 M_{eff} = \mu_0 M_s$, i.e., it is dominated by the shape anisotropy. For the 25 nm-thick sample, the Curie temperature is reduced from the bulk value of 370 K to about 330 K and the saturation magnetization is reduced to 220 mT from its low-temperature bulk value of 500[58] to 220 mT. The 10 nm-thick LSMO top contact layer has a Curie temperature well below 300 K. This is also confirmed by FM resonance measurements which show only one magnetic layer with properties corresponding to the 25 nm bottom layer. The LSMO/PZT/LSMO capacitors are defined using dry etching 80 µm × 80 µm square mesa structures in the top LSMO layer. Electrical contacts to the individual capacitors are made using a probe tip on micromanipulator at the top electrode and a common bottom electrode.

The experiment aims to measure the FM/FE coupling-induced magnetization in the FM layer. In the following we record the rotation of the light polarization upon reflection from the sample as a function of the FE polarization direction. For this, the FE polarization is modulated between up and down directions by applying a square wave voltage signal to the FM/FE capacitor while monitoring the polarization of the reflected light with lock-in detection. A typical result is shown in **Figure 4** (blue points). As one can see, the switching of the FE polarization for applied voltages between 2 and 4 V causes a significant rotation of the light polarization that is comparable in amplitude to the polar Kerr effect signal of the FM LSMO layer (see Figure 3). This implies that we mostly observe the electrooptic Pockels effect induced by the FE polarization and the applied electrical field in the PZT layer.[59] To separate the electrooptic signal and identify a possible FE-controlled magnetic component in the FM layer that is induced by the electronic FE/FM coupling, we apply an additional out-of-plane magnetic field of 600 mT. This field saturates the magnetization in the FM layer along the film normal (as shown in Figure 3) and causes the expected FE/FM coupling-induced magnetization to be mostly in-plane. The corresponding measurement is shown in Figure 4 by the green points. Above the peak-to-peak switching voltage of the FE layer (approximately 3.5 V) we find an offset between both curves of about 1 µrad. According to the polar MOKE amplitude, this implies that FM/FE coupling induces an out-of-plane spin polarization in an interfacial region with a thickness of only 0.5 nm. This finding demonstrates the presence of an electronic contribution to FM/FE coupling and is in line with the very short spin diffusion length expected in a disordered alloy compound such as LSMO.[44] Our observations are also in agreement with earlier electron energy loss spectroscopy experiments with the PZT/LSMO samples where a FE polarization-dependent component of the electronic screening at the interface was identified.[60]

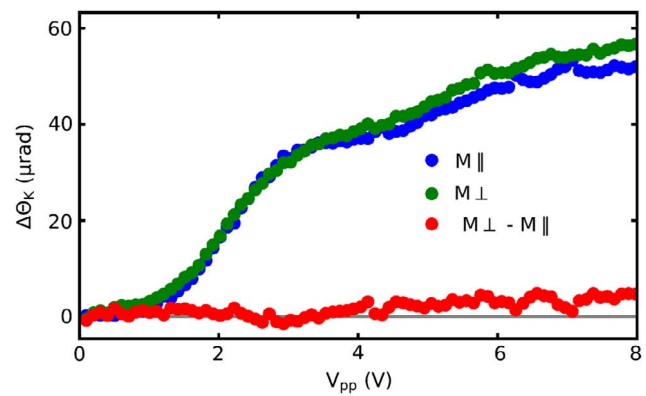

**Figure 4.** Polarization rotation measured in polar geometry as a function of the applied voltage amplitude. For the blue no magnetic field was applied and the FM layer has its magnetization in-plane. The blue curve was recorded with an out-of-plane magnetic field of 600 mT (i.e., out-of-plane magnetization in the FM layer). The red points show the difference between both curves.

## 4. FE Control of Anisotropic Damping in MF Tunnel Junctions

As a further example on the manifestation of the magnonic ME coupling, we present some results on spin-dependent transport in MF tunnel junctions. Such junctions are of relevance for spintronics as a four-point memory element. Our aim here is to discuss the fundamental physics of the process. In principle, the particular dependencies in the tunneling currents shown below can be viewed as the footprints of the magnonic ME coupling, meaning in principle we can use a transport experiment to access the type of ME coupling in these structures.

As mentioned earlier, the interfacial noncollinear order due to the ME coupling results in SOC for the charge carriers. Due to the interfacial origin of the ME coupling, this SOC should be thickness dependent and may show up in the conductance properties of ME tunnel junctions, i.e., a FE layer sandwiched between normal metal (NM) and FM layer. In the presence of the magnonic ME coupling, effects beyond the established tunneling electroresistance[55] can be anticipated.[49] Depending on the orientation of the FE polarization $\vec{P}$, the effective average scattering potential experienced by traversing itinerant electrons from the left to the right and in the opposite direction is different (see **Figure 5**).

In ref. [49], numerical calculations to estimate the strength of the ME coupling effects were conducted. In addition to the conventional Rashba-type SOC at the interface (with strength $\alpha$), we have an additional contribution from the aforementioned ME-induced interfacial magnetic spiral acting on a scale $x_0$. In total, we have then an effective SOC strength $\alpha_R$. As shown in ref. [49], the value of $\alpha_R$ can be related to ME coupling details that we discussed earlier. The magnetic precession can be treated along the lines of the work by Berger.[61] The total number of magnons in FM is given by $n_m = S(1 - \cos\theta_p)n$, where $S$ is the localized spin, $n$ is the number of atoms in FM, and the angle between the magnetic field and the magnetic moment is defined in Figure 5 (left panel). The magnon number satisfies the equation

$$\frac{dn_m}{dt} = -\frac{dn_{\uparrow\downarrow}}{dt} \tag{9}$$

While for the spin-flip rate, we have[30]

$$\frac{dn_{\uparrow\downarrow}}{dt} = \frac{D_N}{4\tau_{\uparrow\downarrow}}(\Delta\mu + \hbar\omega) \tag{10}$$

Here $D_N$ is the electronic density of states in NM, $\Delta\mu$ is the difference of the shifted spin-up and spin-down Fermi level by tunneling current, and $\omega$ is the frequency of the applied microwave field (**Figure 6**).

The electron spin-flip time depends on the spin-wave characteristics and therefore it is space anisotropic, meaning

$$\frac{1}{\tau_{\uparrow\downarrow}} = \left|\frac{d\langle s_{\perp,x}\rangle}{dt}\right|\sin\theta_p \tag{11}$$

The transverse spin component $\langle s_{\perp,x}\rangle$ is averaged with respect to the wave functions which also represent the anisotropy. Typical results of numerical calculations are shown in Figure 2, demonstrating that the magnonic ME coupling can be revealed not only via FMR but also in transport measurements in tunnel junctions. Such measurements can be nowadays also accessed via THz spectroscopy of a traversing THz field, a case which, however, has not yet been analyzed for MF tunnel functions.

## 5. EM Wave-Mediated ME Coupling

### 5.1. Nonlinear MF 1D Chain Models

From a general point of view, the appearance of interfacial spiral magnetic order that we discussed earlier is not surprising. By just imposing the modified boundary conditions on ferromagnetically coupled chain, we end up with a noncollinear order. Thus, one may pose the questions whether it is possible to couple the FE and the FM excitations without directly interfacing FM and FE materials but by subjecting the whole structure to field that imposes certain boundary conditions on the FE and FM ordering. Such an indirect coupling might be mediated by the electric and magnetic field components of an EM wave traversing both the FE and the FM part. To be more specific, let us consider a system consisting of FE/FM/FE/FM/... layers and study the case where the dynamics of the FE and FM modes are decoupled, i.e., no ME coupling is present. Let us now irradiate the structure with an EM wave with appropriate frequency such that FE modes can be excited. The EM wave, meaning both the electric and the magnetic field components of the EM wave, is altered in a specific way depending on the FE response. If the EM wave continues and passes through the FM part with a frequency compatible with the FMR, magnons are excited and the magnetic field is modified. Maxwell's equations imposing strict conditions on the propagation characteristics of the EM wave in space enforce that only certain modes in the FE and the FM part are simultaneously allowed. Therefore, the EM wave in this frequency regime mediates in effect a ME coupling. This scenario can be in its simplest case captured by the following coarse graining model[62,63]

$$H = H_p + H_s$$

$$H_p = \sum_{m=1}^{N/2} \left[ \frac{\alpha_0}{2} \left( \frac{dP_{2m}}{dt} \right)^2 - \frac{\alpha_1}{2} (P_{2m})^2 + \frac{\alpha_2}{4} (P_{2m})^4 - P_{2m} E_{2m}^x \right]$$

$$H_s = -\sum_{m=1}^{N/2} \left[ H_0 S_m^z + D(S_{2m-1}^z)^2 + \vec{H}_{2m-1} \vec{S}_{2m-1} \right]$$

(12)

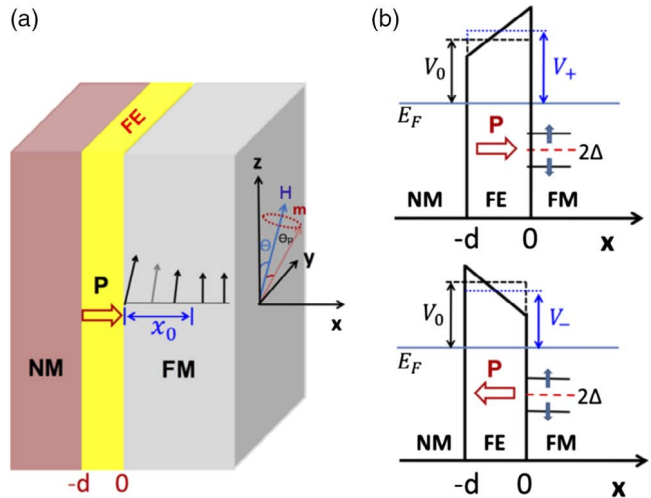

**Figure 5.** Left panel: FE layer sandwiched between NM and FM layers. Right panel: effective potentials experienced by traversing itinerant electrons.

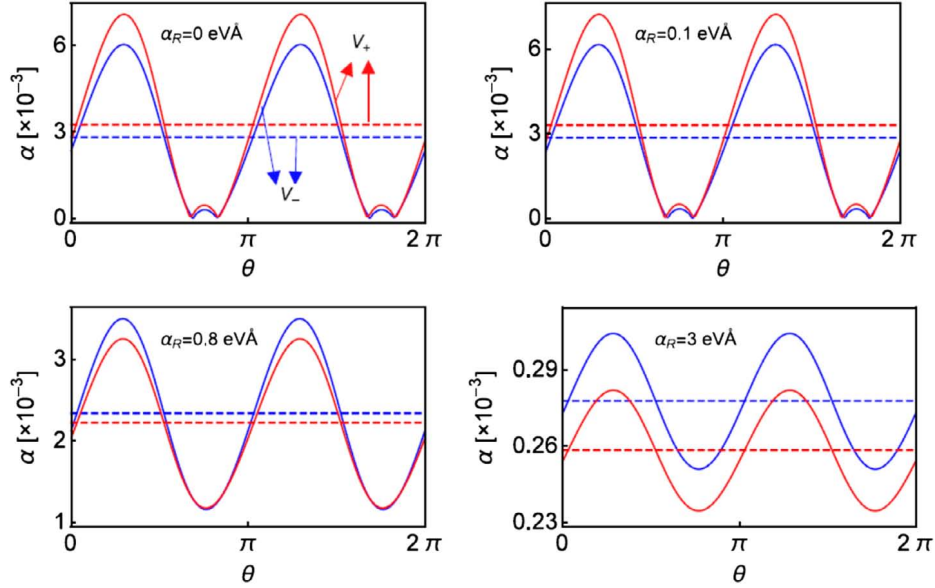

**Figure 6.** Angular dependence of the nonlocal Gilbert damping for different spin-orbit constants $\alpha_R$.

This approach reduces in its linear form to the well-established Maxwell equations with the constitutive relations. The Hamiltonian of the FE layers is $H_p$ and of the FM one is $H_s$ ($P_{2m}$ is the onsite FE polarization and $S_m^z$ is the spin); $\alpha$'s are kinetic coefficients; $E_{2m}^x$ ($\vec{H}_{2m-1}$) is the electric (magnetic) field component of the EM wave in the $2m$ (or $2m-1$) layer; $D$ is a magnetic anisotropy energy; and $H_0$ is a static (global) magnetic field. The consecutive Maxwell equations including the discretized magnetic and electric field components ($h$, $\wp$) have to be solved self-consistently together with the equations of motion for $P_{2m}$ and $S_m^z$. Specifically, one has to solve the following system of coupled equations

$$\frac{1}{c} \frac{d}{dt} (h_{2m+1}^x + 4\pi s_{2m+1}^x) = \frac{1}{2a} (\wp_{2m+2}^y - \wp_{2m}^y)$$

$$- \frac{1}{c} \frac{d}{dt} (h_{2m+1}^y + 4\pi s_{2m+1}^y) = \frac{1}{2a} (\wp_{2m+2}^x - \wp_{2m}^x)$$

$$- \frac{1}{c} \frac{d}{dt} (\wp_{2m}^x + 4\pi p_{2m}) = \frac{1}{2a} (h_{2m+1}^y - h_{2m-1}^y)$$

$$\frac{1}{c} \frac{d\wp_{2m}^y}{dt} = \frac{1}{2a} (h_{2m+1}^x - h_{2m-1}^x)$$

(13)

Here, $p_{2m} = P_0 - P_{2m}$, $\vec{s}_{2m-1} = \vec{S}_0 - \vec{S}_{2m-1}$ are deviations of the FE polarization and magnetization from their equilibrium values; the totality of these deviations constitutes the FE and FM excitations. First, numerical results for a self-consistent solution were presented in ref. [33]. Generally, the aforementioned system of equations allows to describe the following scenario: discretizing the system in the sense of a coarse-grained approach, the spatiotemporal structure of the electromagnetic field at some time $t$ and position $i$ is found from $\vec{D}_i = \epsilon_0(\vec{E}_i + \vec{P}_i)$, $\vec{B}_i = \mu_0(\vec{H}_i + \vec{M}_i)$, where the local electric polarization $\vec{P}_i$ and the magnetization $\vec{M}_i$ are deduced from Equation (13). In this way, nonlinear effects (harmonic generations) and effects of ME coupling are included, which allows to access the ME coupling details by optical means (meaning by analyzing the electromagnetic field vectors after traversing the sample).

## 5.2. Creation and Amplification of Electromagnon Solitons by Electric Field in Nanostructured MFs

Beyond photonic applications, further use of composite FE/FM heterostructures for heat and electromagnetic signal transport is discussed in refs. [39–43]. As an example we discuss here solitonic excitations in such composite structures. Conventional spin waves and solitonic (bullet) waves in FM are well studied; in particular they play a key role in spin-torque nano-oscillators. Our focus is on these type of excitations in the context of MF heterostrucures. Up to now such systems have rarely not been studied in connection with composite MFs. A realization would be similar to the tunnel junction shown in Figure 1 but extended by a giant magnetoresistance (GMR) layer after the first FM layer (which is supposed to act as the free layer) and the GMR layer is followed by another fixed (magnetically hard) FM layer. The oscillators are triggered by current pulses, as shown in Figure 1. Then it would be of interest to inspect the synchronization between two neighboring nanooscillators as a function of an external electric bias that acts on the FE polarization and in doing so we tune the interfacial spin coupling. The basic possibility of coupling between the FE and the FM dynamics as to generate electromagnon solitons was clarified in ref. [46]; the coupling mechanism was assumed to be linear, i.e., the same type as discussed in Section 2. The idea is to trigger, for example, by an electric pulse FE excitations with a frequency in the gap of the (uncoupled) FM system. The ME coupling facilitates the flow of energy into the FM system resulting in electromagnon soliton formation at the interface. An example is shown in **Figure 7**.

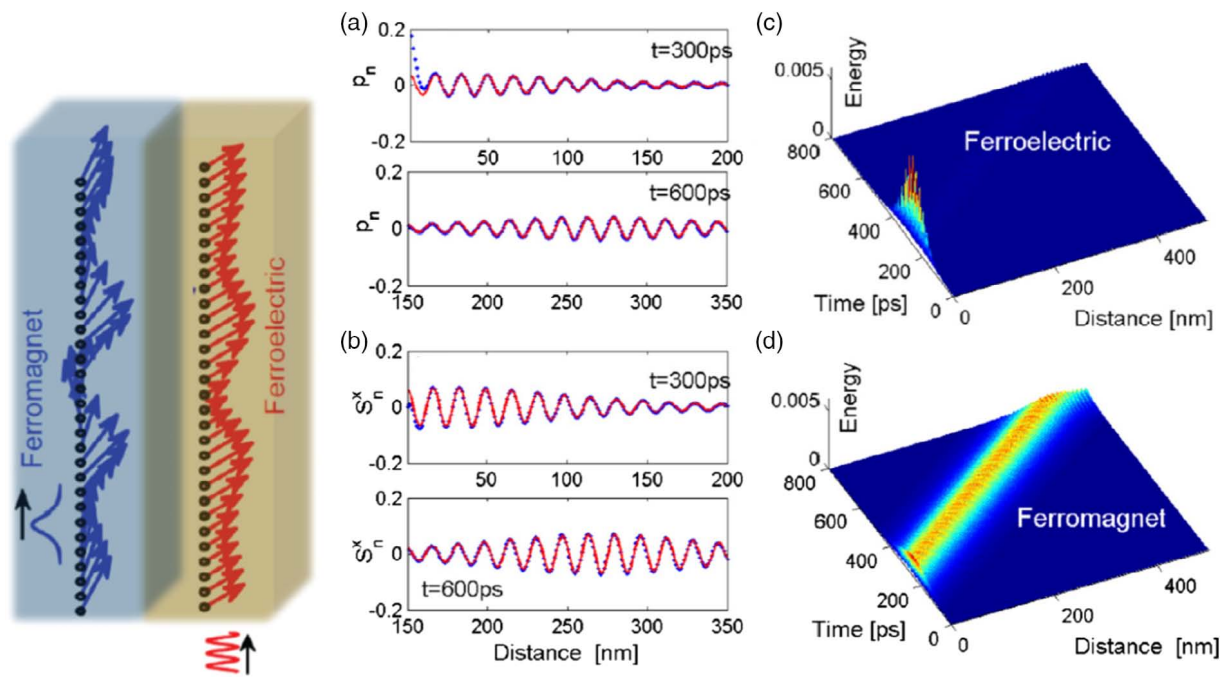

**Figure 7.** (Left) Pictorial representation of coupled FE and FM systems. Arrows indicate the orientation of FE and FM moments. An electric field triggers excitations in the FE order parameter. (Right) Results of numerical simulations: creation of a FM soliton by the electric field and propagation of the electromagnon solitons in the MF system.

For propagation of the electromagnon solitons in the magnetoelectrically coupled layers, it is necessary that the electric field reaches a certain frequency-dependent amplitude, as shown in ref. [64]. This type of composite system is particularly useful for devices where a controllable amplification of the magnonic signal by an electric field is needed.[64] Applications to spin-torque ME coupled nanooscillators and possible applications for E-bias-controlled neuromorphic computing would be highly interesting.

A further important observation concerns Landau–Zener tunneling in the system, as shown in Figure 3, which allows to a controlled FE–FM excitation conversion. Depending on the rise-up time and duration of the external fields, the energy flow between the excitation modes can be controlled, as evidenced by numerical calculations in ref. [65]. Further ongoing studies in this direction are focused on the spatial–temporal control of electron–magnon excitation wave packets (such as in ref. [66]) is nanostructured FE/FM materials.

## 6. Conclusions

In this brief overview we highlighted the delicate noncollinear spin physics that emerges at coupled FE/FM interfaces. We identified an underlying mechanism for ME coupling that is fairly general and relies on the spin-density accumulation at the interface of a dielectric and a FM metal. As the latter is spin polarized, its dielectric screening charge caused by the nearby FE polarization is inevitably spin polarized. The spin-polarized screening density is not in equilibrium with the bulk spin density and therefore it acts with an interfacially localized torque to the bulk magnetization leading so to the formation of a spiral structure at the interface. A number of consequences follow from this interfacial coupling. We gave few examples on its utilization in FE/FM tunnel junctions, in electromagnon solitons, and discuss the photonic properties of FE/FM structures. In view of the vast material compositions and the additional control knobs offered by the ME coupling, it is conceivable that coupled FE/FM systems will continue to attract research from various fields and be part of electrically, magnetically and thermally steered functional devices.

### Acknowledgements

Funding by the DFG through SFB 762, B11, and Mercator fellow (C.J.) is gratefully acknowledged. We thank Simon Wisotzky for assisting in the sample preparation and Marin Alexe and Georg Schmidt for discussions at an early stage of this project.

## Conflict of Interest

The authors declare no conflict of interest.

## Keywords

inhomogeneous electric torque, magnetoelectric coupling, magnetoelectric solitons, two-phase composite multiferroics